\documentclass[12pt]{article}
\usepackage{a4wide,epsfig}

\catcode`@=11
\newcount\@tempcntc
\def\@citex[#1]#2{\if@filesw\immediate\write\@auxout{\string\citation{#2}}\fi
  \@tempcnta\z@\@tempcntb\m@ne\def\@citea{}\@cite{\@for\@citeb:=#2\do
    {\@ifundefined
       {b@\@citeb}{\@citeo\@tempcntb\m@ne\@citea\def\@citea{,}{\bf ?}\@warning
       {Citation `\@citeb' on page \thepage \space undefined}}%
    {\setbox\z@\hbox{\global\@tempcntc0\csname b@\@citeb\endcsname\relax}%
     \ifnum\@tempcntc=\z@ \@citeo\@tempcntb\m@ne
       \@citea\def\@citea{,}\hbox{\csname b@\@citeb\endcsname}%
     \else
      \advance\@tempcntb\@ne
      \ifnum\@tempcntb=\@tempcntc
      \else\advance\@tempcntb\m@ne\@citeo
      \@tempcnta\@tempcntc\@tempcntb\@tempcntc\fi\fi}}\@citeo}{#1}}
\def\@citeo{\ifnum\@tempcnta>\@tempcntb\else\@citea\def\@citea{,}%
  \ifnum\@tempcnta=\@tempcntb\the\@tempcnta\else
   {\advance\@tempcnta\@ne\ifnum\@tempcnta=\@tempcntb \else \def\@citea{--}\fi
    \advance\@tempcnta\m@ne\the\@tempcnta\@citea\the\@tempcntb}\fi\fi}
\catcode`@=12

\begin{document}

\title{\vskip-3cm{\baselineskip14pt
\centerline{\normalsize DESY 02-043\hfill ISSN 0418-9833}
\centerline{\normalsize April 2002\hfill}}
\vskip1.5cm
Heavy Quarkonium Spectrum at ${\cal O}(\alpha_s^5m_q)$
and Bottom/Top Quark Mass Determination.}
\author{ Alexander A. Penin$^{a,b}$,
Matthias  Steinhauser$^a$\\[2mm]
{\small $^a$ II. Institut f\"ur Theoretische Physik, Universit\"at Hamburg,}\\
{\small Luruper Chaussee 149, 22761 Hamburg, Germany}\\
{\small $^b$ Institute for Nuclear Research, Russian Academy of Sciences,}\\
{\small 60th October Anniversary Prospect 7a, Moscow 117312, Russia}}
\date{}

\maketitle

\thispagestyle{empty}

\begin{abstract}
We present the next-to-next-to-next-to-leading 
${\cal O}(\alpha_s^5m_q)$ result for the ground state energy of a
heavy quarkonium system.  On the basis of this  result we  
determine the bottom quark mass from $\Upsilon(1S)$
resonance and provide an explicit formula
relating the top quark mass to the resonance energy 
in $t\bar t$ threshold production.
\medskip

\noindent
PACS numbers: 12.38.Bx, 14.65.Fy, 14.65.Ha
\end{abstract}



\section{Introduction}
Quark masses enter the Lagrangian of Quantum Chromodynamics (QCD) 
as fundamental parameters.
Thus, it is of primary interest to perform accurate determinations of
the mass values combining experimental input with precision
calculations from the theoretical side.
In this letter we focus on the bottom and top quark masses. 
The bottom quark mass is
of particular interest due to the current and future $B$ physics
experiments. In the observables used to extract the
Cabibbo-Kobajashi-Maskawa (CKM) matrix elements and to gain deeper
insight in the understanding of CP violation the bottom quark mass
enters as a crucial input quantity. Thus a precise knowledge of the
latter is essential for the interpretation of the experimental data.
On the other hand, the top quark mass is one of the key parameters for 
precision tests of the standard model of the electroweak 
interactions at high energies and for the search of ``new physics''
at a future $e^+e^-$ linear collider.

Whereas the masses of the up, down and strange quark require
major non-perturbative input the higher quark masses can be 
computed with relatively high accuracy
using perturbative methods.
The study of nonrelativistic heavy-quark-antiquark systems
is one of the most promising approaches in this respect.
These systems allow for a model-independent perturbative treatment
which relies entirely on first principles of QCD \cite{AppPol}.
Nonperturbative effects \cite{Vol,Leu} are well under control for the
top-antitop system and, at least for the ground state resonance, also for
bottomonium. This makes heavy-quark-antiquark systems an ideal laboratory 
to determine the heavy-quark masses. 
Besides its phenomenological importance, the heavy-quarkonium system is very
interesting from the theoretical point of view because it possesses a highly
sophisticated multiscale dynamics and its study demands the full power of the
effective-theory approach. 

Among other characteristics
of the heavy quarkonium system the  ground state energy is of primary
phenomenological interest. It can be directly used to
extract the bottom quark mass from the lowest resonance
of the $\Upsilon$ family
and essentially determines the resonance energy in $t\bar t$
threshold production.  The ${\cal O}(\alpha_s^4m_q)$ 
heavy quarkonium spectrum  was obtained first in \cite{PinYnd} 
and afterwards confirmed in \cite{MelYel2,PenPiv1}. 
At ${\cal O}(\alpha_s^5m_q)$
the heavy quarkonium spectrum was derived in \cite{KPSS2} for 
vanishing $\beta$-function and the large-$\beta_0$ result can be found
in~\cite{KiySum,Hoang:2000fm}. 

In the present letter 
we complete the result of \cite{KPSS2} for the ground 
state energy by calculating the remaining terms proportional to
$\beta$-function. We apply the result to the determination of the
bottom quark mass from the mass of $\Upsilon(1S)$ resonance
and give an explicit formula
relating the top quark mass to the resonance energy 
in $t\bar t$ threshold production both for $e^+e^-$ annihilation and
$\gamma\gamma$ collisions. The main results
are given by Eqs.~(\ref{beta}), (\ref{finalmb1})
and (\ref{ttcorrnum1}).


\section{Perturbative heavy quarkonium ground state energy
at ${\cal O}({\alpha_s^5m_q})$}

For the calculation of the ground state energy 
we follow the general approach of \cite{KPSS2,KPSS1}. It is
based on the nonrelativistic effective theory concept \cite{CasLep,BBL}
in its potential NRQCD (pNRQCD) incarnation \cite{PinSot1}
and is implemented with the threshold expansion \cite{BenSmi}.
In pNRQCD the dynamics of the nonrelativistic potential
heavy quark-antiquark pair 
is governed by the effective  Schr{\"o}dinger equation
and by its multipole interaction to the ultrasoft gluons.
The relativistic effects of the harder modes 
that have been ``integrated out'' are encoded in the 
higher-dimensional operators which appear in the effective Hamiltonian
and correspond  to an expansion in the heavy
quark velocity $v$, and  the  Wilson coefficients
which are series in the strong coupling constant $\alpha_s$.

The analysis of the heavy quarkonium spectrum at 
${\cal O}(\alpha_s^5m_q)$, {\em i.e.} 
the third order corrections to the Coulomb 
approximation involves two basic ingredients:
the effective Hamiltonian and the 
retardation effect associated with the emission and absorption of
dynamical ultrasoft gluons.
The general form of the Hamiltonian valid up to 
next-to-next-to-next-to-leading order\footnote{In 
  such a framework one has two expansion parameters,
  $\alpha_s$ and $v$. The corrections are classified according to the total 
  power of $\alpha_s$ and $v$, {\em i.e.} N$^k$LO corrections contain terms of
  ${\cal O}\left(\alpha_s^lv^m\right)$, with $l+m=k$.
  This has the consequence that, in general, different loop orders, which are
  counted in powers of $\alpha_s$, contribute to the N$^k$LO result.
  Note, that for the corrections to the parameters of a heavy quarkonium
  bound state, where $v\sim\alpha_s$, the standard ordering
  in powers of $\alpha_s$ is restored.} (N$^3$LO) 
is given in~\cite{KPSS2}.
A detailed discussion of the effects 
resulting from the chromoelectric dipole interaction 
of the heavy quark-antiquark  pair to the ultrasoft gluons 
relevant for perturbative bound-state calculations at N$^3$LO
can be found in \cite{KniPen1,KPSS2}.
 
For vanishing angular momentum we can write the 
energy level of the principal quantum number $n$ as
\begin{equation}
E_n^{\rm p.t.}=E^C_n+\delta E^{(1)}_n+\delta E^{(2)}_n+\delta
E^{(3)}_n
+\ldots\,,
\label{enseries}
\end{equation}
where the leading order Coulomb energy is given by 
\begin{eqnarray}
E^C_n&=&-{C_F^2\alpha_s^2m_q\over4n^2}\,.
\end{eqnarray}
$\delta E^{(k)}_n$ stands for corrections of order $\alpha_s^k$.
The first and second order corrections can be found in  
\cite{PinYnd,MelYel2,PenPiv1} for arbitrary $n$.   
For the ground state one gets
\begin{eqnarray}
\delta E^{(1)}_1 &=&E^C_1{\alpha_s\over\pi}
\left[4\beta_0L_\mu+4\beta_0+{a_1\over2}\right],
\nonumber\\
\delta E^{(2)}_1 &=&E^C_1\left({\alpha_s\over\pi}\right)^2
\left[{12}\beta_0^2L_\mu^2+\left(16\beta_0^2
+{3}a_1\beta_0+{4\beta_1}\right)L_\mu
+\left({4}+{2\pi^2\over 3}+{8\zeta(3)}\right)\beta_0^2
\right.
\nonumber\\
&&\left.
+{2a_1\beta_0}+{4\beta_1}+{a_1^2\over 16}+{a_2\over 8}
+\pi^2C_AC_F+\left({21\pi^2\over 16}
-{2\pi^2\over 3}S(S+1)\right)C_F^2\right]\,,
\label{nnlo}
\end{eqnarray}
where $L_\mu=\ln(\mu/(C_F\alpha_sm_q))$,
$\zeta(3)=1.202057\ldots$ is Riemann's $\zeta$-function,
$C_F=4/3$ and $C_A=3$ are the eigenvalues of the quadratic Casimir
operators of the fundamental and adjoint representations of the colour gauge
group, respectively. The modified minimal-subtraction 
($\overline{\rm MS}$) scheme for the renormalization of $\alpha_s$ is implied.
The coefficients $a_i$ parameterize the $i$-loop
perturbative corrections to the Coulomb potential.
The analytical results for the  one- \cite{Fis,Bil} 
and two-loop \cite{Pet,Sch,KPSS1}
coefficients are
\begin{eqnarray}
a_1&=&{31\over 9}C_A-{20\over 9}T_Fn_l\,,
\nonumber \\
a_2&=&\left[{4343\over162}+4\pi^2-{\pi^4\over4}+{22\over3}\zeta(3)\right]C_A^2
-\left[{1798\over81}+{56\over3}\zeta(3)\right]C_AT_Fn_l
\nonumber\\
&&{}-\left[{55\over3}-16\zeta(3)\right]C_FT_Fn_l
+\left({20\over9}T_Fn_l\right)^2\,,
\end{eqnarray}
where $T_F=1/2$ is the index of the fundamental representation and $n_l$ is
the number of light-quark flavors.
For convenience of the reader we also provide
the first three coefficients of the QCD $\beta$-function
\begin{eqnarray}
\beta_0&=&{1\over4}\left({11\over3}C_A-{4\over3}T_Fn_l\right),
\nonumber\\
\beta_1&=&{1\over16}\left({34\over3}C_A^2-{20\over3}C_AT_Fn_l-4C_FT_Fn_l
\right),
\nonumber\\
\beta_2&=&{1\over64}\left({2857\over54}C_A^3-{1415\over27}C_A^2T_Fn_l
-{205\over9}C_AC_FT_Fn_l+2C_F^2T_Fn_l+{158\over27}C_AT_F^2n_l^2
\right.\nonumber\\
&&{}+\left.{44\over9}C_FT_F^2n_l^2\right)
\,.
\end{eqnarray}
The ${\cal O}\left(\alpha_s^3\right)$ corrections 
to the energy levels arise from several sources:
\begin{itemize}
\item[(i)] matrix elements of the N$^3$LO operators of the effective
Hamiltonian between Cou\-lomb wave functions; 
\item[(ii)] higher iterations of the NLO and NNLO operators of the effective
Hamiltonian in time-independent perturbation theory; 
\item[(iii)] matrix elements of the N$^3$LO instantaneous operators generated
by the emission and absorption of ultrasoft gluons; and
\item[(iv)] the retarded ultrasoft contribution.
\end{itemize}

The contributions to $E_n^{(3)}$ for vanishing 
$\beta$-function were computed for general $n$ in \cite{KPSS2}. 
The corresponding correction to the ground state energy
gets contributions from all four sources and reads
\begin{eqnarray}
\left.\delta E^{(3)}_1\right|_{\beta(\alpha_s)=0}
&=&-E^C_1{\alpha_s^3\over\pi}\left\{
-{a_1a_2+a_3\over32\pi^2}
+\left[-{C_AC_F\over2}+\left(-\frac{19}{16}
+{S(S+1)\over2}\right)C_F^2\right]{a_1}\right.
\nonumber\\
&&{}+\left[-{1\over 36}+\frac{\ln2}{6}+\frac{L_{\alpha_s}}{6}\right]
{C_A^3}+\left[-{49\over36}
+{4\over3}\left(\ln2+L_{\alpha_s}\right)
\right]{C_A^2C_F}
\nonumber\\
&&{}+\left[-{5\over 72}+{10\over 3}\ln2
+{37\over6}L_{\alpha_s}+
\left({85\over 54}
-{7\over 6}L_{\alpha_s}\right)S(S+1)\right]
{C_AC_F^2}
\nonumber\\
&&{}+\left[{50\over9}+{8\over3}\ln2
+3L_{\alpha_s}-{S(S+1)\over3}\right]
{C_F^3}
\nonumber\\
&&{}+\left[-{32\over15}+2\ln2+(1-\ln2)S(S+1)\right]{C_F^2T_F}
\nonumber\\
&&{}\left.+{49C_AC_FT_Fn_l\over36}
+\left[{11\over18}-{10\over27}S(S+1)\right]{C_F^2T_Fn_l}
+{2\over3}C_F^3L^E_1\right\}\,,
\label{zerobeta}
\end{eqnarray}
where $S$ is  the spin quantum number and 
$L_{\alpha_s}=-\ln(C_F\alpha_s)$.
The terms proportional to $a_1$ correspond to the iterations of lower-order
operators as mentioned in point (ii) of the above list.
The last term in Eq.~(\ref{zerobeta}) corresponds to 
retarded ultrasoft contribution (cf. point (iv)). 
The corresponding QCD Bethe logarithms can be found in~\cite{KniPen1,KPSS2}
with the numerical result $L^E_1=-81.5379$.
The logarithmic $\ln(\alpha_s)$ part of Eq.~(\ref{zerobeta})
is known from previous analyses \cite{BPSV2,KniPen2}.
At present, only Pad\'e estimates of the three-loop $\overline{\rm MS}$
coefficient $a_3$ are available \cite{ChiEli}. For the bottom and top quark
case they read 
\begin{eqnarray}
{a_3\over 4^3}&=&
\cases{
\displaystyle
\displaystyle
98&if $n_l=4$\,,\cr
\displaystyle
60&if $n_l=5$\,.\cr}
\label{a3}
\end{eqnarray}
Note that at N$^3$LO the Wilson coefficients in the 
effective Hamiltonian are infrared (IR) divergent.
The corresponding divergence  is canceled 
against the ultraviolet (UV) one of the
ultrasoft contribution.
In our analysis  based on the threshold expansion
dimensional regularization is used to handle the divergences.
For convenience, the IR singular pole part of the 
Wilson coefficients is subtracted  according to
the $\overline{\rm MS}$ prescription, so that the coefficients, 
in particular $a_3$, are defined
in the $\overline{\rm MS}$ subtraction scheme both for UV and IR divergences.
For consistency, the UV pole of the ultrasoft contribution is
subtracted in the same way \cite{KPSS2}.

The third order corrections to the ground state energy
proportional to the $\beta$-function 
only get contributions from (i) and (ii).
The computation of the contribution from (i) is rather 
straightforward. It includes the corrections due to the 
three-loop running of the Coulomb potential which we mark
by $C.r.$,
and the one-loop running of the $1/m_q$-suppressed and the
Breit potential $(B.r.)$ of the following form
\begin{eqnarray}
\left.\delta E^{(3)}_1\right|_{C.r.}
&=&E^C_1\left({\alpha_s\over\pi}\right)^3\left\{
{16\beta_0^3}L^3_\mu+\left[{48}\beta_0^3
+{6}a_1\beta_0^2+
{20}\beta_1\beta_0\right]L^2_\mu
\right.
\nonumber \\
&&
+\left[{12\pi^2}\beta_0^3+{12}a_1\beta_0^2
+\left({40}\beta_1+{3\over 4}a_2\right)\beta_0
+{2a_1\beta_1}+{4\beta_2}\right]L_\mu
\nonumber \\
&&
\left.
+{12\pi^2}\beta_0^3+{3\pi^2\over 2}a_1\beta_0^2
+\left({5\pi^2}\beta_1+{3\over 4}a_2\right)\beta_0
+{2a_1\beta_1}+{4\beta_2}
\right\}\,,
\label{Crun}
\\
\left.\delta E^{(3)}_1\right|_{B.r.}
&=&E^C_1{\beta_0\alpha_s^3\over\pi}\left\{
\left[4{C_AC_F}+\left(1
-{4S(S+1)\over 3}\right)C_F^2\right]L_\mu\right.
\nonumber \\
&&
\left.+\left(-1
+{4S(S+1)\over 3}\right)C_F^2\right\}
\,.
\label{Brun}
\end{eqnarray}
The contribution of the type (ii) can be computed using  
the general  approach of \cite{PenPiv1,PenPiv2}  (see also
\cite{KiySum}). 
It includes the corrections due to the 
iterations of the one- and two-loop running of the Coulomb
potential  which we mark by $C.i.$,
and the iteration of the kinetic energy correction,
the $1/m_q$ and the Breit potential with the one-loop running of the Coulomb 
potential $(B.i.)$. The corresponding  analytical expressions
read 
\begin{eqnarray}
\left.\delta E^{(3)}_1\right|_{C.i.}
&=&E^C_1\left({\alpha_s\over\pi}\right)^3\left\{
{16\beta_0^3}L^3_\mu+\left[-{8}\beta_0^3
+{6}a_1\beta_0^2+
{8}\beta_1\beta_0\right]L^2_\mu
\right.
\nonumber \\
&&
+\left[\left(-{20\pi^2\over 3}+64\zeta(3)\right)\beta_0^3
-{2}a_1\beta_0^2
+\left({a_1^2\over 2}+{a_2\over 4}\right)\beta_0
+{a_1\beta_1}\right]L_\mu
\nonumber \\
&&
+\left(-{8}-{8\pi^2}+{2\pi^4\over 45}+64\zeta(3)
-{8\pi^2}\zeta(3)+{96}\zeta(5)\right)\beta_0^3
\nonumber \\
&&\left.
+\left(-{5\pi^2\over 6}+{8\zeta(3)}\right)a_1\beta_0^2
+\left(\left({8}-{8\pi^2\over 3}+{16\zeta(3)}\right)\beta_1
-{a_1^2\over 8}\right)\beta_0
\right\},
\label{Citer}\\
\left.\delta E^{(3)}_1\right|_{B.i.}
&=&E^C_1{\beta_0\alpha_s^3\over\pi}\left\{
\left[4{C_AC_F}+\left(\frac{19}{2}
-{4S(S+1)}\right)C_F^2\right]L_\mu\right.
\nonumber\\
&&\left.+\left({6}-{2\pi^2\over 3}\right){C_AC_F}+
\left(9-{4\pi^2\over 3}
+\left(-{8\over 3}+{4\pi^2\over 9}\right){S(S+1)}\right)C_F^2
\right\}
\,,
\label{Biter}
\end{eqnarray}
where $\zeta(5)=1.036928\ldots$.
As a cross-check to our analytical calculation we 
computed the non-trivial ingredients 
by solving the eigenvalue problem numerically
in  the limit $\alpha_s\to 0$ with the help of the program 
{\tt schroedinger.nb} \cite{Lucha:1998xc}.

By adding (\ref{Crun}), (\ref{Brun}),
(\ref{Citer}) and (\ref{Biter})  we obtain the third order 
correction to the ground state energy proportional to the $\beta$-function
\begin{eqnarray}
\left.\delta E^{(3)}_1\right|_{\beta(\alpha_s)}
&=&E^C_1\left({\alpha_s\over\pi}\right)^3\left\{
{32\beta_0^3}L^3_\mu+\left[{40}\beta_0^3
+{12}a_1\beta_0^2+
{28}\beta_1\beta_0\right]L^2_\mu
\right.
\nonumber \\
&&
+\left[\left({16\pi^2\over 3}+64\zeta(3)\right)\beta_0^3
+{10}a_1\beta_0^2
+\left({40}\beta_1+{a_1^2\over 2}+{a_2}\right.\right.
\nonumber \\
&&
\left.\left.
+8\pi^2C_AC_F+
\left({21\pi^2\over 2}-{16\pi^2\over 3}S(S+1)\right)C_F^2\right)\beta_0
+{3}a_1\beta_1+{4\beta_2}\right]L_\mu
\nonumber \\
&&
+\left(-{8}+{4\pi^2}+{2\pi^4\over45}+64\zeta(3)
-{8\pi^2}\zeta(3)+{96}\zeta(5)\right)\beta_0^3
+\left({2\pi^2\over 3}+{8\zeta(3)}\right)a_1\beta_0^2
\nonumber \\
&&
+\left(
\left({8}+{7\pi^2\over 3}
+{16\zeta(3)}\right)\beta_1
-{a_1^2\over 8}+{3\over 4}a_2
+\left({6\pi^2}-{2\pi^4\over 3}\right)C_AC_F\right.
\nonumber \\
&&
\left.\left.
+\left(8\pi^2-{4\pi^4\over 3}+\left(-{4\pi^2\over 3}
+{4\pi^4\over 9}\right)S(S+1)\right)
C_F^2\right)\beta_0
+{2a_1\beta_1}+{4\beta_2}
\right\}\,.
\label{beta}
\end{eqnarray}
The terms proportional to $\beta_0^3$ in Eqs.~(\ref{beta}) 
can be found in  \cite{KiySum,Hoang:2000fm}.

The total result for the third order correction 
to the ground state energy 
is given by the sum of Eqs.~(\ref{zerobeta}) and (\ref{beta})
\begin{equation}
\delta E^{(3)}_1=\left.\delta E^{(3)}_1\right|_{\beta(\alpha_s)=0}+
\left.\delta E^{(3)}_1\right|_{\beta(\alpha_s)}
\,.
\end{equation}
Adopting the choice $\mu_s=C_F\alpha_s(\mu_s) m_q$ one obtains 
for the bottom and top system (for S=1) in numerical form
\begin{eqnarray}
  \delta E^{(3)}_1&=& \alpha_s^3(\mu_s)E_1^C\left[
  \left(
  \begin{array}{c} 
    70.590|_{n_l=4}\\
    56.732|_{n_l=5} 
  \end{array} 
  \right)
  + 15.297 \ln(\alpha_s(\mu_s)) + 0.001\,a_3 
  + \left(
  \begin{array}{c} 
    34.229|_{n_l=4}\\
    26.654|_{n_l=5} 
  \end{array} 
  \right)\bigg|_{\beta_0^3}
  \right]\,,\quad
  \nonumber\\
  \label{eq:dele3num}
\end{eqnarray}
where we have separated the contributions arising from $a_3$ and
$\beta_0^3$.
The only unknown ingredient in our result for $\delta E^{(3)}_1$
is the three-loop $\overline{\rm MS}$ coefficient $a_3$ 
of the corrections to the static potential entering Eq.~(\ref{zerobeta}).
Up to now there are only estimates based on Pad\'e approximation which
we will use in our analysis.
However, we will show that our final result only changes marginally
even for a rather large deviations of $a_3$ from its  Pad\'e estimate.


\section{Bottom and top quark mass determination}

In this section we focus on the bottom and top 
quark mass determination. We fix the value of the 
strong coupling constant which enters our theoretical expression    
to $\alpha_s^{(5)}(M_Z)=0.1185\pm 0.002$ \cite{PDG}.

{\bf \em Bottom quark.}
The starting point for the determination of the bottom quark mass
is its relation to the mass of the $\Upsilon(1S)$  
resonance
\begin{eqnarray}
  M_{\Upsilon(1S)} = 2m_b + E_1^{\rm p.t.}+ \delta^{\rm n.p.}E_1\,,
  \label{eq:M1SE}
\end{eqnarray}
with $M_{\Upsilon(1S)}=9.46030(26)$~GeV \cite{PDG}.
Here $\delta^{\rm n.p.}E_1$ is the nonperturbative correction
to the ground state energy. 
The dominant nonperturbative correction is associated with the
gluon condensate contribution given by~\cite{Vol,Leu}
\begin{equation}
  \delta^{\rm n.p.}E_1=\frac{1872}{1275}
  {m_q\left\langle\alpha_s G^a_{\mu\nu} G^{a\mu\nu}\right\rangle
  \over (C_F\alpha_sm_q)^4}=60\pm 30~{\rm MeV}\,.
  \label{nptb}
\end{equation}
For the numerical estimate we 
use $\alpha_s^{(4)}(\mu_s)=0.304$ corresponding to the 
soft normalization scale $\mu_s=C_F\alpha_s(\mu_s)m_b$
characteristic to the Coulomb problem and $m_b=4.83$~GeV
which is obtained from Eq.~(\ref{eq:M1SE}) neglecting the 
corrections to the Coulomb binding energy.
This is consistent with neglecting the  perturbative correction
to   Eq.~(\ref{nptb}). 
For the gluon condensate we adopt the value 
from~\cite{Harrison:1998yr}
$\langle \alpha_s G^{a\mu\nu}G^a_{\mu\nu} \rangle = 0.04(2)~\mbox{GeV}^4$.
The effect of  higher dimension condensates on the ground state 
energy is small~\cite{Pineda:1996uk}. 
Note, however, that the contribution of the gluon condensate 
to the energy level of principal quantum number $n$
grows as $n^6$ and becomes unacceptably large already for 
$n=2$. Thus the excited levels are very sensitive
to the nonperturbative long-wave vacuum fluctuations of the 
gluon field and cannot be used for reliable predictions.

\begin{figure}[t]
  \begin{center}
    \begin{tabular}{c}
      \leavevmode
      \epsfxsize=7.5cm
      \epsffile[40 230 550 580]{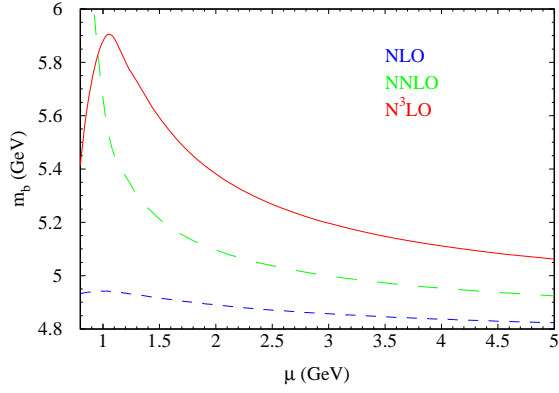}
      \\
      (a)
      \\
      \leavevmode
      \epsfxsize=7.5cm
      \epsffile[40 230 550 580]{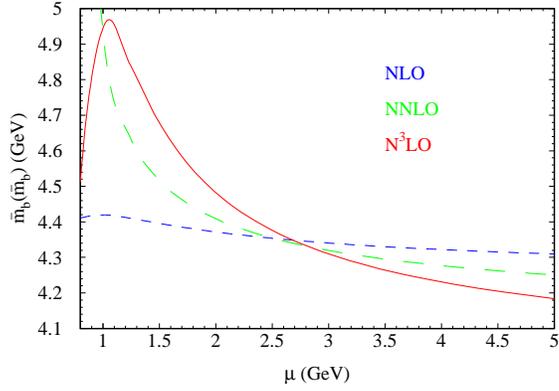}
      \\
      (b)
      \\
      \leavevmode
      \epsfxsize=7.5cm
      \epsffile[40 230 550 580]{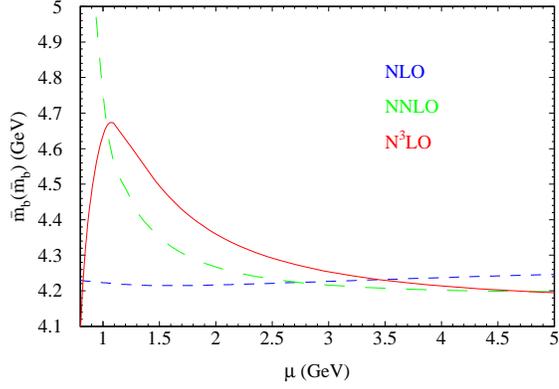}
      \\
      (c)
    \end{tabular}
  \end{center}
  \caption{\label{fig:botpol}
    (a) Bottom quark pole mass, $m_b$, as a function
    of the renormalization scale $\mu$, used for the numerical
    evaluation of $E_1$ where the short-dashed, long-dashed and solid line
    corresponds to the NLO, NNLO and N$^3$LO approximations.
    (b) $\overline{\rm MS}$ bottom quark mass
    $\overline{m}_b(\overline{m}_b)$, 
    as a function
    of the renormalization scale $\mu$, which is used 
    for the extraction of the pole mass $m_b$ (cf. (a)). For the
    conversion from the pole to the $\overline{\rm MS}$ mass the
    method described around Eq.~(\ref{mustar}) is used.
    (c) like (b), however, the upsilon-expansion is used.
          }
\end{figure}

Combining  Eq.~(\ref{eq:M1SE})  with the result of the previous section
for the perturbative ground state energy up to ${\cal O}(m_q\alpha_s^5)$
we obtain the bottom quark mass as a function the renormalization
scale of the strong coupling constant normalization, $\mu$,
which is plotted in Fig.~\ref{fig:botpol}(a).
For the numerical evaluation we extract 
$\alpha_s^{(4)}(m_b)$ with $m_b=4.83$~GeV from its value at $M_Z$ using
four-loop $\beta$-function accompanied with three-loop
matching\footnote{We use the package {\tt
RunDec}~\cite{Chetyrkin:2000yt} to perform the running and
matching of $\alpha_s$}.
$\alpha_s^{(4)}(m_b)$ is used as starting point in order to evaluate
$\alpha_s^{(4)}(\mu)$ at N$^k$LO
with the help of the $k$-loop $\beta$ function.
From Fig.~\ref{fig:botpol}(a) we see that the dependence on the
renormalization scale
becomes very strong below $\mu\sim 2$~GeV which 
indicates that the perturbative corrections are not under control. 
However, even above this scale
the perturbative series for the pole mass shows no sign of convergence.
This means that one can assign a numerical value to the  
pole mass only in a specified order of perturbation theory. 

On the contrary, it is widely believed that the 
$\overline{\rm MS}$ mass ${\overline m}_b(\mu)$
at the scale $\mu={\overline m}_b(\mu)$ is a short-distance object
which has much better perturbative properties. 
Thus, it seems to be reasonable to convert our result for the pole 
mass into ${\overline m}_b({\overline m}_b)$. The relation between $m_b$ 
and ${\overline m}_b({\overline m}_b)$ is known up to 
three-loop approximation
\cite{Chetyrkin:1999ys,Chetyrkin:1999qi,Melnikov:2000qh} and shows
sizable perturbative corrections.
For this reason we suggest the following procedure to take
into account these corrections in a most accurate way.  
The idea is that from the one-parametric family
of ${\overline m}_b(\mu)$
we can choose a representative corresponding to some scale $\mu^\star$ 
in such a way that
\begin{equation}
  {\overline m}_b(\mu^\star)=m_b\,.
  \label{mustar}
\end{equation}
For a given fixed-order value of the pole mass
Eq.~(\ref{mustar}) can be solved for $\mu^\star$.
In particular, for a pole mass to N$^k$LO 
we use the $k$-loop relation between the $\overline{\rm MS}$ 
and pole mass in Eq.~(\ref{mustar}).
Afterwards ${\overline m}_b({\overline m}_b)$ can be computed from
$\overline{m}_b(\mu^\star)$ solving the RG equation.
The advantage of this approach is obvious: we use the finite order relation 
between  $\overline{\rm MS}$ and pole mass at the scale
where they are perturbatively close while the
large difference between ${\overline m}_b({\overline m}_b)$
and $m_b$ is completely covered  by the RG evolution 
which can be computed with very high accuracy as the corresponding  
anomalous dimension is known to four-loop approximation  
\cite{Chetyrkin:1997dh,Vermaseren:1997fq}. The only restriction
on the method could be connected to the value of $\mu^\star$. 
It should be large enough to allow for a reliable use of the RG equation.
As we will see this condition is satisfied in practice. 

In Fig.~\ref{fig:botpol}(b) 
our result for  ${\overline m}_b({\overline m}_b)$  
is plotted at NLO, NNLO and N$^3$LO
as a function of the normalization scale $\mu$ 
which is used to obtain the pole mass $m_b$
(cf. Fig.~\ref{fig:botpol}(a)).
It is remarkable that close to $\mu= 2.7$~GeV, which is consistent
with the physically motivated soft scale $\mu_s\approx 2$~GeV, 
both the second and the third order corrections vanish.
This fact is a rather strong indication of the 
convergence of the series for ${\overline m}_b({\overline m}_b)$.
The numerical values for $m_b$, ${\overline m}_b({\overline m}_b)$ 
at $\mu=2.7$~GeV as well as the values of 
$\mu^\star$ and $\alpha_s^{(4)}(\mu^\star)$ 
are listed in Tab.~\ref{tab:mb} for the different orders. 

An alternative approach to extract ${\overline m}_b({\overline m}_b)$
was suggested in \cite{Hoang:1998hm} 
within the so-called upsilon-expansion.
The basic idea is to correlate the approximation
for the energy spectrum and the one of the 
quark mass relation by matching the leading powers of $\beta_0$.
As a consequence the $k^{\rm th}$-order correction to the spectrum
goes along with the $(k+1)$-loop mass relation.
In this way the large corrections proportional to 
leading powers of $\beta_0$, which are associated to
the IR renormalon contribution, are canceled
in  ${\overline m}_b({\overline m}_b)$.
In order to perform the analysis at N$^3$LO the four-loop
relation between the $\overline{\rm MS}$ and pole mass 
is necessary for which we adopt the
large-$\beta_0$ approximation result~\cite{Beneke:1994sw,Hoang:2000fm}. 
One can expect that it is close to the exact answer as
the large-$\beta_0$ approximation also works well at three-loop
order~\cite{Chetyrkin:1999ys,Chetyrkin:1999qi}.
Adopting this prescription  
we obtain ${\overline m}_b({\overline m}_b)$ as shown in
Fig.~\ref{fig:botpol}(c). 
Note that within the upsilon-expansion the 
``scale of best convergence'' is absent.
Though the NNLO correction vanishes at   $\mu\approx 2.7$~GeV 
as it does within our $\mu^\star$  prescription
the N$^3$LO one at this point is sizable as can be seen from 
Tab.~\ref{tab:mb}. 
This fact can be explained by the observation 
that the hypothesis of the  renormalon contribution 
dominance  in the perturbative series for the energy levels 
does not work at N$^3$LO.
To be precise, if one sets $\mu=\mu_s$ in Eq.~(\ref{beta}) 
one finds that the renormalon
induced term proportional to $\beta_0^3$ 
is indeed large, however, it covers less than a  half of 
the total third order contribution to $E_1$
which can be seen from Eq.~(\ref{eq:dele3num}).
As a consequence, one would have to assign 
a larger uncertainty from the unknown higher order corrections
to the N${}^3$LO value given in the last row of Tab.~\ref{tab:mb}.

\begin{table}
  \begin{center}
    \begin{tabular}{|r||c||c|c|c||c|}
 \hline 
 {order}&
 {$m_b$~(GeV)}&
 {${\overline m_b}({\overline m_b})^{(\star)}$~(GeV)}&
 {$\mu^\star$~(GeV)}&
 {$\alpha_s^{(4)}(\mu^\star)$} &
 {${\overline m_b}({\overline m_b})^{(ups)}$~(GeV)}
 \\
    \hline 
LO&4.749&4.749&---&---&4.324\\
NLO&4.865&4.348&2.498&0.2629&4.223\\
NNLO&5.020&4.338&2.194&0.2816&4.225\\
N$^3$LO&5.235&4.346&1.886&0.3000&4.274\\
    \hline 
    \end{tabular}
    \caption{\label{tab:mb}
    Results for the pole mass $m_b$ and the  
    $\overline{\rm MS}$ mass as computed
    in the $\mu^{\star}$ scheme as described in the text.
    $\mu =2.7$~GeV has been adopted.
    For completeness also $\mu^\star$
    and $\alpha_s^{(4)}(\mu^\star)$ are presented.
    In the last row the values for 
    ${\overline m_b}({\overline m_b})$
    according to upsilon-expansion are given.}
  \end{center}
\end{table}

The uncertainty in the obtained value of 
${\overline m}_b({\overline m}_b)$  originates from
several sources which are listed in Tab.~\ref{tab:mberr}. Beside the  
uncertainties  in $\alpha_s^{(5)}(M_Z)$  and the gluon condensate
one has to take into account the errors in the Pad\'e
estimates of the coefficient $a_3$, the charm quark mass effects 
and the higher order contributions.
Let us consider the corresponding errors in more detail.
In the above analysis we neglected the charm quark
mass both in the correction to the spectrum and 
in $\overline{\rm MS}$-pole mass relation.
As an estimate of the induced error we use the available results
incorporating the charm quark mass effects in the 
spectrum~\cite{Eiras:2000rh,Hoang:2000fm} and
the quark mass relation~\cite{Gray:1990yh,Hoang:2000fm}. 
The value of higher order contributions is  difficult to estimate.
The nice convergence of the series for ${\overline m}_b({\overline m}_b)$ 
in the first three orders implies their small size. 
Commonly the uncertainties from the higher orders are related
to the uncertainty induced by a variation of the normalization
scale $\mu$.
We estimate it by allowing a variation of $\mu$ by $\pm 400$~MeV, 
which corresponds to twice the difference between the 
scales at which  the NNLO and N$^3$LO corrections vanish,
around our optimal value $\mu=2.7$~GeV.  
The uncertainty related to the error in the Pad\'e estimate for $a_3$ 
can be written as $(a_3^{(n_l=4)}/98-1)\times 10$~MeV and 
will completely disappear once the exact computation becomes available.
For the numerical estimate we allow for a $\pm 100\%$ deviation of $a_3$ from 
the Pad\'e estimate. 

Finally, our prediction for the 
bottom quark  $\overline{\rm MS}$ mass reads
\begin{equation}
  {\overline m}_b({\overline m}_b)=4.346\pm 0.070~{\rm GeV}
  \,,
  \label{finalmb1}
\end{equation}
where the uncertainties are added up in quadrature.

\begin{table}
  \begin{center}
\renewcommand{\arraystretch}{1.2}
    \begin{tabular}{|c|r|r|}
    \hline 
    source &  $\delta{\overline m}_b({\overline m}_b)$~(MeV) &
    $\delta E_{\rm res}$~(MeV)\\
    \hline 
   $\alpha_s(M_Z)$                & ${}^{+13}_{-8}$  & ${}^{+149}_{-156}$ \\  
   $\mu$                          & ${}^{+66}_{-47}$ & $\pm 41$  \\
   $a_3$                          & $\pm 10$         & $\pm 22$\\
   $\left\langle GG\right\rangle$ & $\pm 15$         & $-$ \\  
   $m_c$                          & $\pm 10$         & $-$ \\  
   $\delta^{\Gamma_t}E_{\rm res}$ & $-$              & $\pm 10$\\
 \hline
   total                          & $\pm 70$         & $\pm 163$\\
 \hline    
\end{tabular}
\renewcommand{\arraystretch}{1.0}
    \caption{\label{tab:mberr} The sources of uncertainties
    and corresponding errors in the bottom quark 
    $\overline{\rm MS}$ mass
    and  the resonance energy in $t\bar t$ 
    threshold production. The intervals of the input parameters
    variation are specified in the text. The total uncertainty 
    is obtained by adding up the individual ones in quadrature
    where in the case of asymmetric errors the maximum is chosen.}
  \end{center}
\end{table}

Now we are in the position to compare our result
to the estimates obtained with other methods.
There are different approaches to determine 
the bottom quark mass from the
$b\bar b$ pair production close to threshold
which are quite complementary as they 
are based on different experimental and theoretical input.
Three basic techniques can be distinguished:
\begin{itemize}
\item[(i)] the direct analysis of the $\Upsilon$ resonances spectrum;
\item[(ii)]low-order moments  sum rules;
\item[(iii)] the nonrelativistic $\Upsilon$ sum rules.
\end{itemize}
Each approach has its own advantages and shortcomings.  
They are briefly characterized below and the results of the most recent
theoretical analyses based on (i) -- (iii) 
are listed in Tab.~\ref{tab:mbref}.

The direct analysis of the spectrum of $\Upsilon$ resonances~\cite{AppPol} 
used in this letter suffers from large nonperturbative corrections. 
The exited levels  are too sensitive
to the nonperturbative dynamics 
and only the ground state can be used for reliable  predictions.
On the other hand currently the most accurate perturbative
description has been achieved within this framework. 
 
Study of the low-order moments of the spectral density within the sum rules 
approach  relies on the global quark-hadron duality.
Nonperturbative effects are negligible.  
The theoretical analysis consists in fully relativistic
fixed-order calculations  which  have been performed up to NNLO
\cite{Kuhn:2001dm}. The method allows for a direct determination
of the $\overline{\rm MS}$ mass.
It is, however, sensitive to the 
behaviour of the spectral density above the threshold which is not 
well known experimentally.

The  nonrelativistic $\Upsilon$ sum 
rules ~\cite{NSVZ} are ``between'' the approaches (i) and (ii).
They operate with the high (of the order $1/\alpha_s^2$) moments  of the 
spectral density,
which are saturated by the nonrelativistic near-threshold region.
The experimental input is given  by the 
parameters of the  $\Upsilon$ resonances which are known with high accuracy.
On the theoretical side the nonperturbative effects  are well under control,
however, the Coulomb effects should be properly taken into account. 
By now the analysis  has been performed up to 
NNLO~\cite{KPP,PenPiv1,MelYel2,BenSin,Hoang:2000fm}.
The extension to N$^3$LO is a challenging theoretical problem.
In particular, it requires
the calculation of third order corrections to the 
heavy quarkonium wave function at the origin. The
first steps in this direction has been performed 
in~\cite{KPSS2,KniPen1,KniPen2,Pen,KniPen3}.  
 
As we see from the Tab.~\ref{tab:mbref} the results of all
approaches are in reasonable agreement.

\begin{table}
  \begin{center}
    \begin{tabular}{|r|c|c|c|}
    \hline 
    \multicolumn{1}{|c|}{reference}&  \multicolumn{1}{c|}{method} & 
    \multicolumn{1}{c|}{order} & 
    \multicolumn{1}{c|}{${\overline m}_b({\overline m}_b)$~(GeV)} \\
    \hline 
    PP98 \cite{PenPiv1} & nonrelativistic $\Upsilon$ sum rules &  NNLO &  
    $4.21 \pm 0.11$ \\  
    MY99 \cite{MelYel2} & " & " &$4.20 \pm 0.10$\\  
    BS99 \cite{BenSin} & " & " & $4.25\pm 0.08$\\  
    H00   \cite{Hoang:2000fm} & " & " & $4.17 \pm 0.05$\\
    KS01  \cite{Kuhn:2001dm} &  low moments sum rules &  " &
    $4.191 \pm 0.051$  \\
    P01   \cite{Pineda:2001zq}  & spectrum, $\Upsilon(1S)$ resonance & "  & 
    $4.210 \pm 0.090\pm0.025$ \\
    this work  & " & N$^3$LO  &  $4.346 \pm 0.070$ \\
    \hline 
    \end{tabular}
    \caption{\label{tab:mbref}  Recent results for 
    the  $\overline{\rm MS}$ mass obtained in different approaches 
    and approximations.}
  \end{center}
\end{table}

{\bf\em Top quark.} 
Let us now turn to the top quark case where the experimental
data are not yet available. Our goal is to present
a formula which can be directly used for the top quark mass 
determination from the characteristics of the 
cross section of $t\bar t$ production near threshold.
One has to distinguish the production
in  $e^+e^-$ annihilation
where the final state quark-antiquark pair is produced
with $S=1$ and in (unpolarized) $\gamma\gamma$
collisions where the dominant contribution is given by the
$S$-wave zero spin final state. 

The nonperturbative effects in the case of the top quark are
negligible, however, the effect of the
top quark width has to be taken into account properly~\cite{FadKho}
as the relatively large width smears out the Coulomb-like resonances
below the threshold. 
The NNLO analysis of the cross section~\cite{hoaetal}
shows that only the ground state pole results in the well-pronounced 
resonance.
The higher poles and continuum, however, affect the position of the 
resonance peak  and  move it to higher energy.
As a consequence, the  resonance energy can be written
in the form
\begin{eqnarray}
  E_{\rm res}&=&2m_t+E_1^{\rm p.t.}+\delta^{\Gamma_t}E_{\rm res}\,.
  \label{ttcorr}
\end{eqnarray}
To compute the shift of the peak position 
due to the nonzero top quark width $\delta^{\Gamma_t} E_{\rm res}$
we use the result of \cite{PenPiv2} for the $e^+e^-\to t\bar t$
and $\gamma\gamma\to t\bar t$ cross sections 
at NNLO. Numerically we get for both processes
\begin{equation}
  \delta^{\Gamma_t} E_{\rm res}=100\pm 10~{\rm MeV}\,.
  \label{engam}
\end{equation}
This value is rather stable with respect to the variation of 
all input parameters of our analysis such as the top quark
mass and width, the value of  the strong coupling constant
and its renormalization scale. 
In the following we adopt as our central input value
$\Gamma_t=1.43$~GeV~\cite{PDG} and use for the 
soft renormalization scale $\mu_s=C_F\alpha_s(\mu_s)m_t$.
For $m_t=174.3$~GeV~\cite{PDG} the latter gives $\mu_s\approx 30$~GeV.

\begin{figure}[t]
  \begin{center}
    \begin{tabular}{c}
      \leavevmode
      \epsfxsize=8cm
      \epsffile[40 230 550 580]{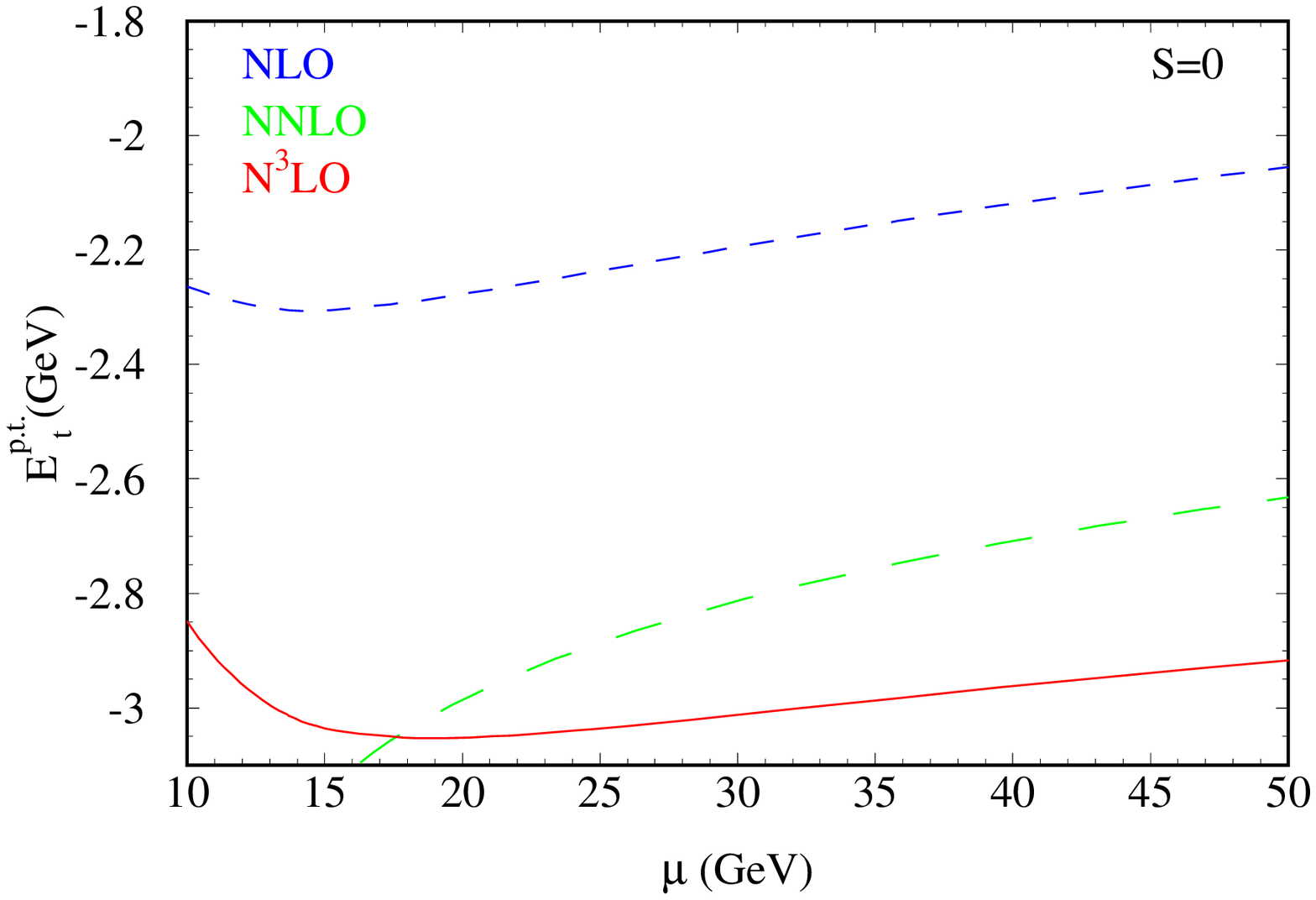}
      \\
      (a) \\
      \leavevmode
      \epsfxsize=8cm
      \epsffile[40 230 550 580]{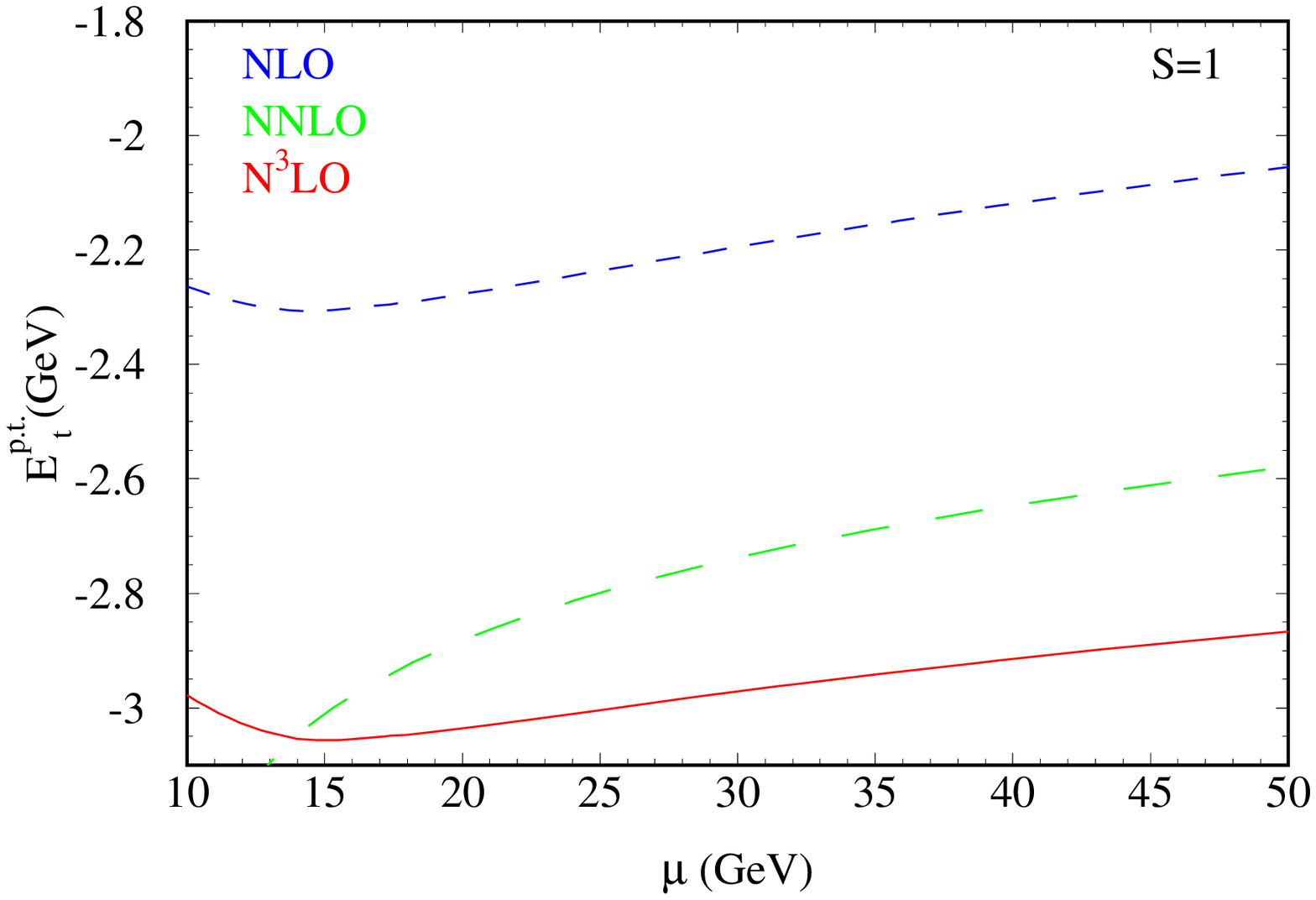}
      \\ (b)
    \end{tabular}
  \end{center}
  \caption{\label{fig:top1}
  Ground state energy $E^{\rm p.t.}_1$ of the 
  $t\bar{t}$ bound state in the zero-width approximation 
  as a function of the renormalization scale $\mu$
  for $S=0$ (a) and $S=1$ (b). 
  The short-dashed, long-dashed and solid line
  corresponds to the NLO, NNLO and N$^3$LO approximations.
          }
\end{figure}

To evaluate  the perturbative contribution we use
the result of the previous section up to N$^3$LO. 
In Figs.~\ref{fig:top1}(a) and (b) the perturbative
ground state energy  in NLO, NNLO and N$^3$LO is 
plotted as a function of the renormalization scale 
of the strong coupling constant
for $S=0$ and $S=1$, respectively.
One can see, that the N$^3$LO result shows a much weaker 
dependence on $\mu$ than the NNLO one. 
Moreover at the scale $\mu\approx 15$~GeV for $S=1$
and   $\mu\approx 18$~GeV for $S=0$,
which are  close to the physically motivated scale $\mu_s$,
the N$^3$LO correction vanishes and furthermore 
becomes independent of $\mu$, {\it i.e.} the N$^3$LO curve shows a local
minimum. This suggests
the convergence of the  series for 
the ground state energy {\em in the pole mass scheme}, which 
is inconsistent with the renormalon picture. 
Transforming the result to the $\overline{\rm MS}$ top quark
mass leads to a worse convergence of the series for the resonance
energy. However, the difference in the series behaviour due to the use of
different mass parameters is less significant than in the case of the bottom
quark as $\alpha_s$ for the top system is much smaller 
(see, {\it e.g.},~\cite{Chetyrkin:1999qi}).
Note that the result shows very weak spin dependence.

Collecting all the contributions  we obtain a universal 
relation between the resonance energy in $t\bar t$ threshold 
production in $e^+e^-$ annihilation or $\gamma\gamma$ collisions
and the top quark pole mass  
\begin{equation}
  E_{\rm res}= (1.9833 
  + 0.007\,\frac{m_t-174.3~\mbox{GeV}}{174.3~\mbox{GeV}}
  \pm 0.0009)\times m_t\,,
  \label{ttcorrnum1}
\end{equation}
where the weak nonlinear dependence of
Eq.~(\ref{ttcorr}) on $m_t$
is taken into account in the second term on the right-hand side of
Eq.~(\ref{ttcorrnum1}).
The central value is computed for $m_t=174.3$~GeV.
The sources of the theoretical uncertainty in 
Eq.~(\ref{ttcorrnum1}) are listed in Tab.~\ref{tab:mberr}
in terms of the resonance energy.
For the estimates
we  assume a $\pm 100\%$ error in the  Pad\'e approximation  
of the coefficient $a_3$ and vary the normalization scale
in the stability interval which is roughly given by
$0.4\mu_s<\mu<\mu_s$. 
Due to the very nice behaviour of the perturbative expansion
for the ground state energy we do not expect large higher 
order corrections to our result. 


\section{Conclusions}

In this letter we have completed the result of \cite{KPSS2} for the
N$^3$LO corrections to the 
ground state energy of perturbative heavy quarkonium by computing
the contributions proportional to the $\beta$-function, Eq.~(\ref{beta}).
We have found that the  
N$^3$LO corrections are dominated neither by logarithmically
enhanced $\alpha_s^3\ln(\alpha_s)$ nor by the 
renormalon induced  $\beta_0^3\alpha_s^3$ terms 
(cf. Eq.~(\ref{eq:dele3num}))
and thus the full calculation of the correction is crucial for 
quantitative analysis.

The  result has been applied  to 
determine the bottom quark mass from the mass of the $\Upsilon(1S)$
resonance.  The perturbative series for the extracted bottom quark pole 
mass diverges 
and should be removed from the analysis  in favour of
the short-distance mass 
${\overline m}_b({\overline m}_b)$ to get a reliable prediction.
Our result~(\ref{finalmb1}) 
for ${\overline m}_b({\overline m}_b)$  is in reasonable agreement 
with the NNLO estimates obtained in the framework of nonrelativistic 
$\Upsilon$ sum rules and the approach based on low moments, however,
with a slightly higher central value.

The explicit formula~(\ref{ttcorrnum1})
relating  the top quark mass to the resonance energy 
in $t\bar t$ threshold production both in 
$e^+e^-$ annihilation and $\gamma\gamma$ collisions has been derived. 
In contrast to the bottom quark case  the perturbative expansion
for the ground state energy of a (hypothetical)
top-antitop bound state converges very well in the pole mass scheme. 
As a consequence the top quark pole mass
can be extracted from the future experimental data 
on $t\bar t$ threshold production with the theoretical uncertainty 
of $\pm 80$~MeV.
 
\bigskip
\noindent
{\bf Acknowledgments}
\smallskip

\noindent
We thank B.A. Kniehl and V.A. Smirnov for carefully reading the 
manuscript and useful comments and A. Hoang for discussions about the
upsilon-expansion.
This work was supported in part by the Deutsche Forschungsgemeinschaft through
Grant No.\ KN~365/1-1 and by the Bundesministerium f\"ur Bildung und Forschung
through Grant No.\ 05 HT1GUA/4.

\end{document}